\documentclass{article} 
\usepackage{custom,times}

\usepackage{amsmath,amsfonts,bm}

\def\eqref#1{equation~\ref{#1}}

\def\1{\bm{1}}

\def\ra{{\textnormal{a}}}

\def\rx{{\textnormal{x}}}

\def\rva{{\mathbf{a}}}

\def\erva{{\textnormal{a}}}

\def\ervx{{\textnormal{x}}}

\def\rmA{{\mathbf{A}}}

\def\vmu{{\bm{\mu}}}
\def\vtheta{{\bm{\theta}}}
\def\va{{\bm{a}}}

\def\ve{{\bm{e}}}

\def\vx{{\bm{x}}}

\def\eva{{a}}

\def\mA{{\bm{A}}}

\def\mH{{\bm{H}}}
\def\mI{{\bm{I}}}
\def\mJ{{\bm{J}}}

\def\mX{{\bm{X}}}

\def\mSigma{{\bm{\Sigma}}}

\DeclareMathAlphabet{\mathsfit}{\encodingdefault}{\sfdefault}{m}{sl}
\SetMathAlphabet{\mathsfit}{bold}{\encodingdefault}{\sfdefault}{bx}{n}
\newcommand{\tens}[1]{\bm{\mathsfit{#1}}}
\def\tA{{\tens{A}}}

\def\tX{{\tens{X}}}

\def\gG{{\mathcal{G}}}

\def\sA{{\mathbb{A}}}
\def\sB{{\mathbb{B}}}

\def\sS{{\mathbb{S}}}

\def\emA{{A}}

\newcommand{\etens}[1]{\mathsfit{#1}}

\def\etA{{\etens{A}}}

\newcommand{\E}{\mathbb{E}}

\newcommand{\R}{\mathbb{R}}

\newcommand{\KL}{D_{\mathrm{KL}}}
\newcommand{\Var}{\mathrm{Var}}

\newcommand{\Cov}{\mathrm{Cov}}

\newcommand{\normltwo}{L^2}
\newcommand{\normlp}{L^p}

\newcommand{\parents}{Pa} 

\DeclareMathOperator*{\argmax}{arg\,max}
\DeclareMathOperator*{\argmin}{arg\,min}

\DeclareMathOperator{\Tr}{Tr}

\usepackage{hyperref}
\usepackage{url}
\usepackage{graphicx}
\usepackage{amsmath}
\usepackage{xspace}

\def\sysname{\textsc{Adder}\xspace}
\def\sysnamec{\textsc{Adder2}\xspace}
\def\ada{\textsc{OpenAI Ada}\xspace}
\def\oai{\textsc{oai}\xspace}
\newtheorem{example}{Example}
\newtheorem{remark}{Remark}
\def\qs{{\mathbb{Q}}}
\def\q{{q}}
\def\cs{{\mathbb{C}}}
\def\similar{{\mathtt{sim}}}
\def\retoai{{\mathtt{ret}_{\mathtt{oai}}}}
\def\retadr{{\mathtt{ret}_{\mathtt{adr}}}}
\def\retadrt{{\mathtt{ret}_{\mathtt{adr2}}}}
\def\hardneg{{\mathtt{GlobalNeg}}}

\def\smcalflow{{\textsc{SMCalFlow}}\xspace}
\def\nltox{{\textsc{NL2X}}\xspace}
\def\nltosmcalflow{{\textsc{NL2SMCalFlow}}\xspace}
\def\bash{{\textsc{Bash}}\xspace}
\def\nltobash{{\textsc{NL2Bash}}\xspace}
\def\python{{\textsc{Python}}\xspace}
\def\nltopython{{\textsc{NL2Python}}\xspace}
\def\conala{{\textsc{Conala}}\xspace}
\def\stackoverflow{{\textsc{StackOverFlow}}\xspace}

\def\ds{{\mathbb{D}}}
\def\c{{c}}
\def\E{{E}}
\def\L{{\mathbb{L}}}
\def\Tr{{\mathtt{Tr}}}
\newcommand\ignore[1]{}
\DeclareMathOperator*{\argmaxk}{\mathtt{arg}\,\mathtt{maxK}}

\title{Augmented Embeddings for Custom Retrievals}

\author{Anirudh Khatry, Yasharth Bajpai, Priyanshu Gupta, Sumit Gulwani \& Ashish Tiwari \\
Microsoft Corp.\\
Redmond, WA 98052, USA\\
\texttt{\{t-akhatry, ybajpai, priyansgupta, sumitg, astiwar\}@microsoft.com} 
}

\iclrfinalcopy 
\begin{document}

\maketitle

\begin{abstract}
Information retrieval involves selecting artifacts from a corpus that are most relevant to a given search query.  The flavor of retrieval typically used in classical applications can be termed as {\em{homogeneous}} and {\em{relaxed}}, where queries and corpus elements are both natural language (NL) utterances (homogeneous) and the goal is to pick most relevant elements from the corpus in the Top-K, where K is large, such as 10, 25, 50 or even 100 (relaxed).  Recently, retrieval is being used extensively in preparing prompts for large language models (LLMs) to enable LLMs to perform targeted tasks.  These new applications of retrieval are often {\em{heterogeneous}} and {\em{strict}} -- the queries and the corpus contain different kinds of entities, such as NL and code, and there is a need for improving retrieval at Top-K for small values of K, such as K=1 or 3 or 5.  Current dense retrieval techniques based on pretrained embeddings provide a general-purpose and powerful approach for retrieval, but they are oblivious to task-specific notions of similarity of heterogeneous artifacts.  We introduce {\em{Adapted Dense Retrieval}}, a mechanism to transform embeddings to enable improved task-specific, heterogeneous and strict retrieval. Adapted Dense Retrieval works by learning a low-rank residual adaptation of the pretrained black-box embedding.  We empirically validate our approach by showing improvements over the state-of-the-art general-purpose embeddings-based baseline.
\end{abstract}

\ignore{
\begin{abstract}
Information retrieval refers to identifying an artifact from a large corpus that is most relevant for a given search query.
Retrieval is widely used, most recently for preparing prompts for large language models (LLMs) as part of the retrieval augmented generation paradigm.
A canonical way for retrieving is based on using general-purpose embeddings provided by language models, whereby cosine similarity on the
embeddings defines a distance metric and retrieval is performed by finding nearest neighbors of the query in the corpus.
This general approach for retrieval is powerful, but it is oblivious to any task-specific notion of similarity of artifacts.
We introduce Feature Attention, a mechanism to transform embeddings to enable improved task-specific retrieval where we learn to attend to different features represented in the embeddings differently for a given specific down-stream task. 
We empirically validate our approach by showing improvements over the state-of-the-art general-purpose embeddings-based baseline.
\end{abstract}
\endignore}

\section{Introduction}

Information retrieval has a long and diverse history.
A variety of approaches have been 
proposed~\citep{inferenceIR,probabilisticIR,rankingIR,bertIR,seIR},
yet retrieval continues to remain a challenging problem.
The goal of retrieval is simple: extract the most relevant artifacts from a large corpus
given a query. Retrieval approaches are broadly classified as sparse or dense. 
Sparse retrieval refers to approaches that are based on sparse
representations of the query and corpus, such as a bag of words 
representation~\citep{ir}. Sparse retrieval suffers from 
vocabulary mismatch problem, which is solved by dense retrieval.
Dense retrieval exploits dense vector representations, or embeddings, 
of the queries and corpus elements and uses them to compute the
similarity between query and corpus elements. Hybrid approaches combine the
two by using sparse methods first to select promising candidates and then
dense methods to pick from those candidates~\citep{nogueira2019multistage}.
%

In dense retrieval, the vector representations of the query and the corpus elements
play a very important role in determining the quality of 
retrieval. Models that compute these vector representations can be pretrained.
Pretrained word embeddings~\citep{mikolov2013efficient} and 
sentence embeddings~\citep{reimers-2019-sentence-bert} are widely used.
The pretraining happens on unsupervised data at scale and helps create high
quality vector representations of text and even code~\citep{neelakantan2022text};
for example, the \ada embedding model is one such pretrained model.

There is renewed interest in retrievals and embeddings due to the wide adoption
of generative pretrained large language models (LLMs), such as {\texttt{GPT-4}}~\citep{openai2023gpt4}.
These models have demonstrated the remarkable emergent ability of performing
new tasks when provided only a few examples of the task~\citep{NEURIPS2020_1457c0d6}.
The models' performance on new tasks also improves when they are provided with
information relevant to the task in the prompt. Searching and collecting all information
that may be relevant to a task is a {\em{retrieval}} problem, which leads to
the retrieval-augmented generation (RAG) paradigm~\citep{lewis2020retrieval}.
A popular approach to perform retrieval for RAG is to use 
dense retrieval; that is, find embeddings for the query and corpus and
then use cosine similarity  to pick the best artifacts for the query.
However, the performance of dense retrieval for RAG applications has often
been found to be lacking.
This is because the retrieval problems that arise in RAG have certain features that are different
from the more traditional applications of retrieval.

\begin{figure}
\center
\includegraphics*[scale=1.5]{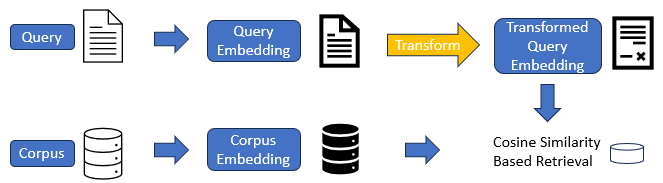}
\caption{Overview of adapted dense retrieval: The embedding of the query is additionally transformed before doing a cosine similarity with embedding of the corpus element. In the full variant, the corpus embedding is also transformed (not shown in figure) before performing cosine similarity.}\label{fig:overview}
\end{figure}

\begin{enumerate}
    \item First, these retrieval problems are {\em{heterogeneous}} in nature because the corpus here may contain code, documentation, structured and semi-structured text, data, and artifacts in specialized domain-specific languages. The embeddings of these artifacts created by pretrained models (such as \ada)
may not necessarily be high quality because the pretraining data may not have had access
to those kinds of special-purpose artifacts. 
\item Second, the retrieval results are {\em{strictly}} required to contain all
relevant artifacts in the top-$k$ for much smaller $k$. This is because the prompts sent to LLMs can
include a limited amount of information (tokens) due to cost and model-induced constraints.
So, here top-10 or top-50 accuracy matters much less than the top-1 or top-5 performance.
\item Finally, the retrieval problems are more {\em{nuanced}} due to task-specific notions
of what is similar or relevant and what is not. This notion of similarity may not match
the notion of similarity learnt by general-purpose embeddings produced by pretrained
models.
\end{enumerate}
The situation with pretrained embeddings is similar to the situation with pretrained
language models. Pretrained generative language models~\citep{gpt,gpt2,gpt3} perform well on several tasks,
but there are always some specialized domain-specific tasks where their performance
is poor. One approach that has been used to address this issue is 
{\em{fine tuning}}~\citep{wortsman2022robust,liu2022fewshot}.
Fine tuning achieves transfer learning by further training the weights of a pretrained model 
on new task-specific data. In the same spirit, we ask the question:
{\em{can we fine tune pretrained embedding models with task-specific data?}}
This can potentially solve the three challenges mentioned above. However, fine tuning requires
examples and access to weights of a model. We often have access to training data, but only in the form
of a {\em{limited number}} of examples of ``good'' retrievals. Moreover, we often do not have 
access to the weights of the model, and even if we did, the cost and data requirements would be
prohibitive. We solve both these issues by so-called 
{\em{parameter efficient fine tuning}} (PEFT)~\citep{liu2022fewshot}.
We instantiate PEFT in our setting by augmenting the black-box embedding model with small ``adapters''.
The new parameters 
introduced in the adapters are learnt using the limited amount of training data.
The internal weights of the black-box embedding model are left untouched.
The adapted embeddings are then used for retrieval. We call our approach
{\em{Adapted Dense Retrieval}} or \sysname; see Figure~\ref{fig:overview}.
The \sysname approach is based on tuning  the
embedding models and this allows us to effectively support heterogeneity,
strictness requirements, and semantic alignment with task-specific notions of similarity.

The adapter we attach to the black-box model in \sysname takes the form of a 
residual adaptation term. Intuitively, the adapter
is learning the {\em{perturbation}} to be applied to the general-purpose embedding
so that it becomes more task and domain specific. 
This is reminiscent of low-rank adaption (LoRA),
which has been used to fine tune large language models by adding
low rank matrices to the weight matrices of the original model~\citep{hu2021lora}.
The low rank matrices can be viewed as perturbations.
Our residual adaption is different since it does not change any weights,
but rather it adds a perturbation to the embedding itself.
We compute the perturbation term by passing the general-purpose embedding through a
softmax based dictionary lookup. 
Mathematically, the dictionary lookup 
looks like a low rank matrix multiplication but for the fact that there is softmax
layer in between the two matrices in our case.

Once the architecture for the adapter is finalized, the final step is to use the training data
that contains examples of ``good'' retrievals to learn the adapter weights.
Let us say we have a training data point where a query $\q$ maps to a corpus element $\c$.
We train our adapter by enforcing that the 
adapted embedding of $\q$ be closer to the adapted embedding of $\c$
than to the adapted embedding of any negative corpus element $\c'$.
A key question here is what negative example $\c'$ to use when processing the
data point $(\q,\c)$.
%
%
There are several works that describe how to pick the negative training instances for dense 
retrieval~\citep{karpukhin-etal-2020-dense}. We take cue from the 
work~\citep{xiong2020approximate} that suggested taking hard negatives from the full corpus.
Specifically, instead of random or in-batch local negatives, we find global negatives using the 
current state of the adapted model on the entire corpus.
%



\section{Heterogeneous Strict Retrieval}
\label{sec:dr}

Retrieval is being increasingly used to build applications that are based on pretrained large language models~(LLMs). Embeddings from LLMs~\cite{openaiemb, touvron2023llama} have been used extensively to support the retrieval component in such applications. However, some of these new applications of retrieval pose different set of challenges from what were posed by other more traditional uses of retrieval. Motivated by the new applications, we introduce the class of {\em{heterogeneous strict retrieval}} problems where we are given three inputs:

\vspace{-0.2cm}
\begin{itemize}
    \setlength\itemsep{0.1em}
    \item A single query $\q_0$ 
      from a domain $\qs$ of queries;
    \item A corpus $\cs$ of candidate items;
    \item A dataset $\ds$ containing pairs $(\q,\c)$, where $\q\in \qs$ and $\c\in \cs$.
\end{itemize}
\vspace{-0.2cm}

The goal is to extract some $k > 0$ elements from $\cs$ that are most likely to be related
to the query $\q_0$ in the same way that pairs in $\ds$ are related.
The focus of our work is on cases where $k$ is required to be small. We use the term 
{\em{strict}} retrieval to emphasize the fact that we want the most likely candidate to appear 
in the top-1 or top-3 or top-5 of the retrieved candidates (as opposed to being in the top-10
or top-50).

In the applications we are targeting, 
the corpus $\cs$ typically contains a few hundreds or a few thousands of elements.
Similarly, the dataset $\ds$ contains a few hundreds or thousands of $(\q,\c)$ pairs.
A key distinction with most earlier work on retrieval is that we are targeting 
{\em{heterogeneous}} retrievals where the domains $\qs$ and $\cs$ are potentially very distinct.
For example, the artifacts in $\qs$ may be
natural language sentences, whereas those in $\cs$ may be code snippets, or semi-structured 
or structured text or data.

We assume that we have black-box access to a pretrained large language model that can compute
embeddings. Let $\E: (\qs\cup \cs) \mapsto \L$ be the black-box function that given an entity --
either a query $\q$ from $\qs$ or a corpus element $\c$ from $\cs$ -- computes its representation
$\E(\q)$ or $\E(\c)$, respectively, in some (common) latent representation 
space $\L$.
For our experiments, we will use the text embedding model \ada as the function $\E$. 
We can use the model to compute (general-purpose) embeddings, but we assume we have no ability to fine-tune or
refine the model in any way. This assumption is reasonable since access to model weights is
often restricted, and furthermore, fine-tuning the models can be cost and time prohibitive.
Moreover, we typically have only small amounts of data in set $\ds$ that may not be sufficient for
fine tuning.

\subsubsection*{Dense Retrieval}

We first recall dense retrieval, which serves as a baseline for solving
our {\em{heterogeneous strict retrieval}} problem.

The embeddings computed by pretrained large models are {\em{dense}} (as opposed to sparse),
and such embeddings can be used for retrieval 
using the following function,
$\retoai(k, \q_0, \cs)$, which 
retrieves $k$ candidates 
 from the corpus $\cs$ for a given query $\q_0$:
\begin{eqnarray}
 \retoai(k, \q_0, \cs) & ::= & \argmaxk_{\c\in \cs}\;\;  \similar( \E(\q_0), \E(\c) )  \label{eqn:oai}
\end{eqnarray}
where $\similar$ is a measure of similarity between two 
vectors in the embedding space $\L$~\citep{xiong2020approximate,DBLP:conf/acl/LeeCT19,luan-etal-2021-sparse,karpukhin-etal-2020-dense}.
We use cosine similarity as the measure in this work; thus,
the $\argmaxk$ operator is simply returning the $k$ corpus elements that 
are the $k$ nearest neighbors of $\E(\q_0)$ in the latent space $\L$.

The embeddings computed by $\E$ are general-purpose and can be used as latent representations
for various entities, such as, code and natural language. The function
$\retoai$ defined in Equation~\ref{eqn:oai} provides a very strong baseline for retrieval, even
for {{\em{heterogeneous strict retrieval}},
since large language models have been trained on vast amounts of data. 
However, it relies on a fixed notion of similarity, which is not informed by 
the specific notion of similarity that may be of interest. In our setting, we are given the
set $\ds$ of pairs of query and corpus elements that illustrates the notion of 
similarity that is desired. 
If the similarity notion implicit in $\ds$ matches the one
induced by $\E$ and $\similar$, then dense retrieval, as performed by $\retoai$,
would perform just as well. However, if there is a mismatch,
then we need to {\em{adapt}} the embedding so that it captures the semantics used in 
the relation between the example pairs in $\ds$.

{\bf{Remark.}} Note that we are using the same embedding calculator $\E$ for both query and corpus in Equation~\ref{eqn:oai} only for simplicity of presentation.
We can generalize easily to the case when we have different functions, say $\E$ and $\E'$, to embed queries and corpus
elements respectively as long as the target latent space $\L$ is shared. All subsequent developments also generalize naturally.

\ignore{
[From the ANCE paper]
Task Definition: Given a query $q$ and a corpus $C$, the (first stage)
 retrieval is to find a set of documents
relevant to the query $D^+ = \{d_1, \ldots, d_i, \ldots, d_n\}$ from $C$ ($D+ \subseteq C$), which then serve as input to
later more complex models (Croft et al., 2009). Instead of using sparse term matches and inverted
index, Dense Retrieval calculates the retrieval score $f()$ using similarities in a learned embedding
space 
(Lee et al., 2019; Luan et al., 2020; Karpukhin et al., 2020):
$$f(q, d) = sim(g(q; \theta), g(d; \theta)),$$  
where $g()$ is the representation model that encodes the query or document to dense embeddings.
The encoder parameter $\theta$ provides the main capacity. The similarity function ($sim()$) is often simply
cosine or dot product to leverage efficient ANN retrieval (Johnson et al., 2017; Guo et al., 2020).
\endignore}

\section{Adapted Dense Retrieval}
\label{sec:adr}

Our approach for addressing the {{\em{heterogeneous strict retrieval}} challenge is based on
adapting the black-box embedding $\E$ to the specific notion of similarity presented
in the dataset $\ds$. We call our approach {\em{adapted dense retrieval}}. 

We adapt the embedding function $\E$ by simply composing it with another function
$\Tr: \L \mapsto \L$ so that the initial query $\q_0$ is now transformed into
$\Tr(\E(\q_0))$. The transformation $\Tr$ is learnt from the dataset $\ds$, which we will describe later.
Thus, the dense retrieval function $\retoai$ from Equation~\ref{eqn:oai} is replaced
now by the following {\em{adapted dense retrival}}, or \sysname, function 
$\retadr(k, \q_0, \cs)$:
\begin{eqnarray}
 \retadr(k, \q_0, \cs) & ::= & \argmaxk_{\c\in \cs}\;\;  \similar( \Tr(\E(\q_0)), \E(\c) )  \label{eqn:adr}
\end{eqnarray}

The \sysname~function is illustrated in Figure~\ref{fig:overview}.
The reader may notice the asymmetry in Equation~\ref{eqn:adr} wherein the transformation
$\Tr$ is applied only the embedding of the query. We can also apply another transformation $\Tr'$
on the corpus element to get a variant 
$\retadrt(k, \q_0, \cs)$
of adapted dense retriever that we call an
{\sysnamec}:  
\begin{eqnarray}
 \retadrt(k, \q_0, \cs) & ::= & \argmaxk_{\c\in \cs}\;\;  \similar( \Tr(\E(\q_0)), \Tr'(\E(\c)) )  \label{eqn:adrt}
\end{eqnarray}
Thus, our \sysname~approach retrieves $k$ nearest neighbors of the transformed embedding of the query.
The purpose of the transformations $\Tr$ and $\Tr'$ is to capture any task-specific adaptations 
required on the embedding before it is used for nearest neighbor search.

The overall approach of adapted dense retrieval can work with transformations $\Tr$ and $\Tr'$ 
taken from any rich class of transformations. In this paper, we start with very simple
classes of transformation for picking $\Tr$ and $\Tr'$, which is described next.

\subsection{Key-Value Lookup based Residual Adaptations}

Inspired by prior work on low-rank adaptations that were used for fine-tuning large language
models~\citep{hu2021lora}, we use residual adaptation as the transformation
function $\Tr$ and $\Tr'$.
Specifically, the two transformations add a residual term to the embedding $e$ computed by the
black-box model; that is,
\begin{eqnarray}
 \Tr(e) & ::= & e + f(e, \theta) \nonumber
\\
 \Tr'(e) & ::= & e + f'(e, \theta') \label{eqn:tr}
\end{eqnarray}
The functions $f$ and $f'$ are parameterized by $\theta$ and $\theta'$ respectively that are
learned from data. The neural architecture for computing $f$ and $f'$ is inspired
by key-value dictionary lookup in an attention mechanism. 

Since $f'$ is handled the same way as $f$, let us just focus on $f$.
The function $f$ is computed by extrapolating from a table that contains mapping from
keys (inputs of $f$) to values (outputs of $f$). Concretely,
let $d$ be the dimension of the latent space $\L$ where the embeddings computed by the
black-box model reside. For example, for OpenAI ada embeddings, $d$ is $1536$.
Let $h < d$ be a small number compared to $d$. The typical values we use for $h$ range from $16$ to $128$. 
Let $K$ be a $(h\times d)$-matrix of keys (containing $h$ keys), and
let $V$ be a $(h\times d)$-matrix of values (containing $h$ values).
The matrices $K,V$ define the unknown parameters $\theta$, and hence
given the input embedding $e$ (of shape $(1\times d)$),
the function $f$ is then defined as:
\begin{eqnarray}
 f(e, K, V) & := & \mathtt{softmax}(e K^T) V
\end{eqnarray}
where $K^T$ is the transpose of $K$. 
Note that since $h < d$, 
even though the final residual term is $d$ dimensional, 
all residuals technically lie in some smaller $h$ dimensional space.

Our low-dimensional residuals are reminiscent of 
low-rank adaption from~\citet{hu2021lora}. However, unlike that work, our 
approach treats the model as a black-box and we do not
adapt the weights of the model.

We have thus parameterized $f$ (and $f'$) by $2dh$ unknowns, and we next describe the loss
function that can help us learn the unknown parameters from the given dataset $\ds$ containing
pairs $(\q,\c)$ of query $\q$ and corpus element $\c$.

\subsection{Global Negatives Contrastive Loss}

Finally, we need to learn the parameters $K,V$, or $\theta$ in short, for the transformation $\Tr$ to obtain
our adaptive dense retrievers outlined in Equation~\ref{eqn:adr} and 
additionally parameters $K',V'$, or $\theta'$ in short, 
for $\Tr'$ to obtain the retriever in Equation~\ref{eqn:adrt}.

The dataset $\ds$ contains pairs $(\q,\c)$, so for each $\q$ in the dataset, we have 
examples of corpus elements $\c$ that should be retrieved.
Every other corpus element becomes a negative example; that is,
for query $\q$, the positive corpus elements $\cs^{\q+}$ and the negative elements
$\cs^{\q-}$ are defined as follows:
\begin{eqnarray}
  \cs^{\q+} & ::= & \{\c\in\cs \mid (\q,\c)\in\ds\}
\\
  \cs^{\q-} & ::= &  \cs \setminus \cs^{\q+}
\end{eqnarray}

Let $g(\q,\c)$ be the following function that is used in our adapted retriever
and that is computed by our neural network: 
\begin{eqnarray}
 g(\q, \c) & ::= & \similar( \Tr(\E(\q)), \Tr'(\E(\c)) )  \label{eqn:g}
\end{eqnarray}
Learning to retrieve is the same as learning to rank~\citep{10.1561/1500000016},
and hence,
to learn the best possible $g$ for retrieval, we need to minimize the loss
over all pairs of positive and negative samples for a query; that is, 
\begin{eqnarray}
 \theta^*,{\theta'}^* & ::= & \argmin_{\theta,\theta'} 
  \sum_{(\q,\c)\in\ds, \c^-\in\cs^{\q-}} l(\Tr(\E(\q)), \Tr'(\E(\c)), \Tr'(\E(\c^-)) )  \label{eqn:kv}
\end{eqnarray}
If $e := \Tr(\E(\q))$, $e^+ := \Tr'(\E(\c))$, and $e^- := \Tr'(\E(\c^-))$, then the
loss function $l(e,e^+,e^-)$ should be zero if $e$ is closer to $e^+$ than it is to $e^-$ and it should be positive otherwise. 
Since the set $\cs^{\q-}$ can be very large, often one approximates by sampling from this set and using
the sample in Equation~\ref{eqn:kv} in place of $\cs^{\q-}$. Typically, the sample is restricted to the corpus elements
that appear in the batch being processed. However, recent work showed that this leads to poor learning~\citep{xiong2020approximate}.
That work also emphasized the need for finding negative samples from the full corpus. We observed the same behavior, and hence we picked the negative corpus element that was closest to the query embedding as {\em{the}} negative sample to use in Equation~\ref{eqn:kv}.
In other words, we use the following modification of Equation~\ref{eqn:kv}:
\begin{eqnarray}
 \theta^*,{\theta'}^* & ::= & \argmin_{\theta,\theta'} \;
  \sum_{(\q,\c)\in\ds} l(\Tr(\E(\q)), \Tr'(\E(\c)), \Tr'(\E( \hardneg(\q))) )  \label{eqn:kv1}
\end{eqnarray}
where the function $\hardneg$ is computed as follows:
\begin{eqnarray}
 \hardneg(\q) & ::= & \argmax_{\c\in \cs^{\q-}}\;\; \similar( \Tr(\E(\q)), \Tr'(\E(\c)) ) \label{eqn:kv2}
\end{eqnarray}
Note that the ideal negative sample, $\hardneg(\q)$, for query $\q$ depends on the functions $\Tr$ and $\Tr'$ that we
are learning. Thus, $\hardneg(\q)$ has to be calculated during training in every batch.
This is not an issue since we have relatively small training datasets and modest corpus sizes.


\ignore{ 
\subsubsection*{Old Text: Applications}
Embedding transformer for few-shot selection:
Let us say we learn the embedding transformer EF that maps an Embedding(NL) to Embedding(M-programs). 
When we get a new NL query, q, we compute EF(Embedding(q)) and then 
use it to find closest M-code in our database for few-shot selection. 

\begin{example}
Here is an example of what we gain by using embedding transformers over just comparing query with NL. 
Note that "remove rows" in Pandas is the same as "select rows" if you look at the code, but if you look at NL, the 
word "remove" is opposite of "select". So, when we pick NL queries using our new approach, we pick few shots 
where we are illustrating selection of rows.

****By Code similarity Query: remove rows where Defense is 100
- 0.9065214395523071: remove rows where Defense is 100
- 0.8953106999397278: get all 'HP' entries where the value is '> 50'
- 0.8910788297653198: get rows in which HP <50 and >100
- 0.8906436562538147: Drop infinite values from 'Attack' and 'Defense' in dataframe
- 0.8835802674293518: remove rows where Total is less than Negative
****By NL similarity Query: remove rows where Defense is 100
- 0.9999999403953552: remove rows where Defense is 100
- 0.778232216835022: drop rows that are duplicates in both 'HP' and 'Attack'
- 0.7573599219322205: Drop infinite values from 'Attack' and 'Defense' in dataframe
- 0.7568846940994263: Remove column positive
- 0.751843273639679: remove rows where Total is less than Negative
\end{example}

Embedding transformer for reranking:
If LLM generates N candidates, we take cosine similarity of Embedding(candidate) with EF(Embedding(q)) to rerank the N candidates.

\begin{remark}
Instead of just learning a embedding transformation function $F_{nl}$ such that 
$Ei_{Code}. F_{nl}(Ei_{NL})$ is (close to) 1, we can conjointly learn embedding transformation 
functions $F_{code}$ and $F_{NL}$ such that $F_{code}(Ei_{Code}) . F_{nl}(Ei_{NL})$ is (close to) 1. 
\end{remark}

\begin{remark}
This can be generalised to domain specific code - code or nl-nl retrievals. For instance if we talk about semantic retrieval for the code fix scenario, we want to match the current code snippet and error message context to code snippets that maybe "relevant" for fixing it. Learning an embedding Transformer each for both query and keys can help us learn transformations in a new "code fix space" in which features relevant for the downstream task are appropriately scaled/combined
\end{remark}

\begin{remark}
The transformation functions will perform a non-linear transform  to  $E_{nl}$ and $E_{code}$ to $E_{code-nl}$ that brings out the relevant feature from both the embedding space for the nl-code mapping we desire. 
\end{remark}
\endignore}

\section{Experimental Evaluation/Results}

We next experimentally evaluate {\em{adapted dense retrievals}} in different ways.
The goal is to determine if adaptation adds any value over off-the-shelf \ada embeddings.

When we use the approach from Equation~\ref{eqn:adr}, we refer to the system
as $\sysname$. When we use Equation~\ref{eqn:adrt}, the system is called $\sysnamec$.
The main baseline is $\oai$, which denotes the use of standard \ada embeddings ({\em{text-embedding-ada-002}}\footnote{https://openai.com/blog/new-and-improved-embedding-model}).
We carried out our experiments on regular laptops and desktops, and used no special purpose
hardware for training or inference except for the black-box rest API calls to the OpenAI embedding
endpoint.

{\em{Experimental Setup:}} To be precise, for our experiments over \sysname~\& \sysnamec, we use a virtual machine with a single Nvidia K80 GPU (with 24GiB of vRAM), 4 CPU cores and 28 GB of RAM.
The optimization is done using the Adam optimizer~\citep{kingma2014adam} for both mechanisms. For training of \sysname, we start an initial learning rate of $1e^{-3}$ with a step update of $\gamma=0.5$ every 100 epochs. On the other hand, for \sysnamec, we start with a learning rate of $1e^{-2}$ and a step update with $\gamma=0.5$ every 50 epochs. The training for all cases was terminated after 500 epochs (although learning saturated in most cases well within the terminal limit). The best model weights are determined using hyperparameter tuning in a range of hidden layer dimensions between 64 and 1536.

\subsection{\sysname\ improves retrieval on certain IR benchmarks}

We first evaluate adapted dense retrieval on classical retrieval benchmarks.
We use benchmarks from the BEIR collection~\citep{thakur2021beir}.
In particular, we focus on four datasets from the BEIR collection: 
SciFact, ArguAna, FiQA and NFCorpus.
These were the four smallest sets in the BEIR collection, see Table~\ref{table:datasets}, and thus they aligned with our motivation
of focusing on tasks where there was limited data. They are also a reasonably diverse set of
heterogeneous retrieval use cases.
The dev test was used to train the model in case of NFCorpus. ArguAna contains only a test set, so we did a 80-20 split to 
create train-test partitions. \oai\ needs no training for retrieval, but 
\sysname\ and \sysnamec\ were trained on the particular training sets, and all were
evaluated on the test sets. 

\begin{table}[t]
\caption{Metadata for BEIR benchmarks used for evaluation.}
\label{table:datasets}
\begin{center}
\begin{tabular}{l|cccc}
Name & Partitions & Queries Used & Corpus size 
\\ \hline
SciFact & train, test &  919 train, 339 test & 5K
\\
ArguAna & test &  1406 test split into 1124 train,282 test & 8.67K
\\
FiQA & train, dev, test & 14166 train, 1706 test & 57K 
\\
NFCorpus & train, dev, test & 11385 dev as train, 12334 test & 3.6K
\\ \hline
\end{tabular}
\end{center}
\end{table}

\begin{table}[t]
 \resizebox{0.96\textwidth}{!}{\begin{minipage}{\textwidth}

\caption{Retrieval Performance on BEIR Datasets. The metrics for SBert were uniformly worse than those for OpenAI embeddings, and hence they are not reported here. Boldface denotes a win for the tool in that category and underline denotes a tie.}
\label{table:beir}
\begin{center}
\begin{tabular}{l|cccc|cccc|cccc}
& 
\multicolumn{4}{c|}{\oai\ nDCG@} & 
\multicolumn{4}{c|}{\sysname\ nDCG@} & 
\multicolumn{4}{c}{\sysnamec\ nDCG@} 
\\ \hline 
 Benchmark
 & 1&3&5&10 
 & 1&3&5&10 
 & 1&3&5&10 
\\ \hline
SciFact 
 & 0.62&0.69&0.70&0.73
 & 0.65&0.71&0.77&0.77
 & {\bf{0.74}}&{\bf{0.80}}&{\bf{0.81}}&{\bf{0.83}}
\\
ArguAna 
 & 0.32&0.49& {\bf{0.54}}&{\bf{0.59}}
 & {\bf{0.37}}&{\bf{0.49}}&0.53&0.57
 & 0.36&0.48&0.52&0.56
\\
FiQA
 & {\bf{0.41}}&{\bf{0.39}}&{\bf{0.40}}&{\bf{0.43}}
 & 0.40&0.37&0.39&0.41
 & 0.39&0.37&0.38&0.41
\\
NFCorpus 
 & 0.46&0.42&{\underline{0.40}}&{\underline{0.37}}
 & {\bf{0.47}}&{\bf{0.43}}&{\underline{0.40}}&{\underline{0.37}}
 & 0.45&0.40&0.37&0.35
\\ \hline
\end{tabular}
\end{center}
\end{minipage}
}
\end{table}

Table~\ref{table:beir} presents the standard ``normalized discounted cumulative gain'' (nDCG) values
at $k=1, k=3, k=5,$ and $k=10$ for the $4$ benchmarks and the $3$ systems.
SciFact and ArguAna datasets associate only one corpus element with any query. 
Hence, for these two cases, nDCG@$k$ can only increase with $k$.
That is not true for the FiQA and NFCorpus.

The systems \sysname\ and \sysnamec\ both perform better than \oai\ on SciFact and ArguAna
in terms of the nDCG metrics @1, 3, 5. 
For FiQA and NFCorpus, the adapted retrivers, \sysname\ and \sysnamec, are unable to
add much value over \oai\ and mostly perform almost as good as \oai.

The results show that adapted dense retrieval can improve performance of black-box
pretrained embedding models by learning a transformation that is customized to the
particular retrieval task. This tuning of the pretrained embedding does not need 
huge amounts of training data. Task-specific adaptation helps when the
train set contains enough examples of retrieving the various artifacts in the corpus.
In cases where it does not improve performance metrics, adaptation does not make the
metrics much worse and performance remains almost at par with the black-box embedding model.
We hypothesize that the differences in gains achieved by 
\sysname and \sysnamec across benchmarks
is due to the differences in alignment of the notion of semantic similarity learnt by 
\ada with what is intended in the benchmarks. For example, SciFact uses very stylized queries
and the notion of similarity is that the query is either supported by or refuted by the corpus
text. The adapter is possibly able to tweak the embeddings to match the benchmark intention for
this set. The notion of similarity is closer to NL semantic similarity for NFCorpus, so there is 
little opportunity to gain by adaptation.


\ignore{
RQ1 : Do the learned embedding transformers retrieve elements similar in the target domain?
Ie is the predicted embedding indeed close to the true embedding of the true target.

RQ2 : Can embedding transformers be used to improve performance on downstream tasks?

RQ3 : Is it robust to choice of model we use to compute the embedding?

I learnt a NL-Embedding to Code-Embedding transformer using train set. Then on the test set, I used it to predict the code-embedding for a given NL query. Then I ranked the LLM generated candidates using cosine similarity with the predicted code embedding. Sherry gave me some LLM-generated candidates but they were already ranked by her non-execution-based-interleaving-technique -- which is already a nontrivial technique. Each candidate was annotated by whether it was a "execution match" (with ground truth). Our technique performed as well as the nontrivial baseline. In some cases, it even gave a 4\% to 6\% improvement over it (but in some cases it also degraded by the same amount depending on the train-test split). There were only 144 benchmarks in this experiment. 

The second experiment was using embedding transformer for predicting which columns will be used in the generated code. So, I learnt a NL-Embedding to Column-name-embedding. Using it to select columns (cosine similarity between column-name-embedding and the column-name-embedding-predicted-by-our-transformer) performed the same as using cosine similarity of column-name embedding with NL-embedding. But using the transformer was more robust to perturbations than the other. Robust meaning that handling changes to the NL description that adds some text or redundant information. Again, promising result, but not convincing enough.
\ignore}

\ignore{ 
\begin{table}[t]
\caption{Retrieval Performance on BEIR Datasets. The metrics for SBert were uniformly worse than those for OpenAI embeddings, and hence they are not reported here.}
\label{table:beir}
\begin{center}
\begin{tabular}{lcccccccccccc}
& 
\multicolumn{3}{c}{nDCG@1} & 
\multicolumn{3}{c}{nDCG@3} & 
\multicolumn{3}{c}{nDCG@5} & 
\multicolumn{3}{c}{nDCG@10}
\\ \hline 
 & \oai & \sysname & \sysnamec 
 & \oai & \sysname & \sysnamec
 & \oai & \sysname & \sysnamec
 & \oai & \sysname & \sysnamec
\\ \hline \\
scifact 
 & 0.62 & 0.65 & 0.74  
 & 0.69 & 0.71 & 0.80  
 & 0.70 & 0.77 & 0.81  
 & 0.73 & 0.76 & 0.83  
\\
arguana 
 & 0.31 & 0.37 & 0.36  
 & 0.48 & 0.49 & 0.48  
 & 0.53 & 0.53 & 0.52  
 & 0.59 & 0.57 & 0.56  
\\
fiqa
 & 0.41 & 0.40 & 0.39  
 & 0.39 & 0.37 & 0.37  
 & 0.40 & 0.39 & 0.38  
 & 0.43 & 0.41 & 0.41  
\\
nfcorpus 
 & 0.46 & 0.47 & 0.45  
 & 0.42 & 0.43 & 0.40  
 & 0.40 & 0.40 & 0.37  
 & 0.37 & 0.37 & 0.35  
\\
\end{tabular}
\end{center}
\end{table}
\endignore}

\subsection{\sysname\ significantly improves NL2X retrievals}

We next evaluate the adapted dense retrieval on heterogeneous tasks where
the query is in natural language, but the corpus elements are code fragments.
Our hypothesis is that adaptation will help more in these cases since the
query and corpus elements are different kinds of entities. While the BEIR 
benchmarks were also heterogeneous, the query and corpus were both
natural language expressions.
A second interesting goal here is to study the impact of the popularity of the
programming language on the performance of baseline embeddings and the
adapted embeddings.

\begin{table}[t]
 \resizebox{0.92\textwidth}{!}{\begin{minipage}{\textwidth}

\caption{NL2X Retrieval Metrics}
\label{table:nl2x}
\begin{center}
\begin{tabular}{l|cccc|cccc|cccc}
& 
\multicolumn{4}{c}{\oai\ nDCG@} & 
\multicolumn{4}{c}{\sysname\ nDCG@} & 
\multicolumn{4}{c}{\sysnamec\ nDCG@} 
\\ \hline 
 X 
 & 1 & 3 & 5 & 10
 & 1 & 3 & 5 & 10
 & 1 & 3 & 5 & 10
\\ \hline
\smcalflow 
 & 0.64 & 0.72 & 0.74 & 0.75
 & {\bf{0.96}} & {\bf{0.98}} & {\bf{0.98}} & \underline{0.98}
 & 0.95 & 0.97 & 0.97 & \underline{0.98}
\\
\bash 
 & 0.75 & 0.84 & 0.85 & 0.86
 & 0.78 & \underline{0.88} & \underline{0.89} & \underline{0.89}
 & {\bf{0.79}} & {\underline{0.88}} & {\underline{0.89}} & {\underline{0.89}}
\\
\conala 
 & {\bf{0.77}} & {\bf{0.87}} & {\bf{0.88}} & \underline{0.88}
 & 0.75 & 0.85 & 0.86 & 0.86 
 & 0.75 & 0.86 & 0.87 & \underline{0.88}
\\ \hline
\end{tabular}
\end{center}
\end{minipage}
}
\end{table}

We picked three \nltox\ datasets from the public domain. The three
target languages X we picked were \smcalflow, \bash, and \python.
These language cover the spectrum from low resource, special-purpose,
and less-known language, \smcalflow, on one end to popular, general-purpose
and widely-used language, \python, on the other end with \bash\ somewhere
in the middle.
The \nltosmcalflow benchmark\citep{SMDataflow2020,smcalflow} consists of
user utterances (in natural language) about tasks involving
calendars, weather, places, or people, 
which is paired with an executable dataflow program written in lisp-like
syntax.
The \nltobash benchmark~\citep{nl2bash} consists of natural description of 
a task paired with a bash command that accomplishes that task.
The \nltopython benchmark~\cite{conala} comes from data collected from
StackOverflow, and contains pairs of natural language and aligned code (in \python).
Each benchmark has a well-defined train and test split provided.

The three \nltox\ benchmarks were originally created for the task of generating
code from natural language. Here, we repurpose them for retrieval: given the
natural language utterance and the corpus of program expressions, retrieve the
code corresponding to the NL utterance.
We remark here that this retrieval task is motivated by few-shot selection.
In code generation from NL tasks, one has to pick those NL-Code pairs as few-shots where the code is most similar
to the code expected from the NL query. Any progress on our repurposed task will facilitate few-shot selection.

As before, we can solve this retrieval task using \ada\ embeddings (\oai), or we could
use our approach and adapt the embeddings using the train set and then use them
for retrieval (\sysname\ and \sysnamec).
Table~\ref{table:nl2x} presents the results of retrieving code from the test set
given the corresponding NL utternace from the test set. For \nltosmcalflow,
\sysname\ and \sysnamec\ both give a huge gain
over \oai\ in the nDCG@$k$ metric for all values of $k$.
The gains are still there for \nltobash\ benchmark, but they are a bit muted.
The gains disappear for \nltopython (\conala), and \oai\ starts out-performing
 our models, but only by a slight margin, for small values of $k$.

\ignore{  
\begin{table}[t]
\caption{NL2X Retrieval Metrics}
\label{table:nl2x}
\begin{center}
\begin{tabular}{lcccccccccccc}
& 
\multicolumn{3}{c}{nDCG@1} & 
\multicolumn{3}{c}{nDCG@3} & 
\multicolumn{3}{c}{nDCG@5} & 
\multicolumn{3}{c}{nDCG@10}
\\ \hline 
 & \oai & \sysname & \sysnamec 
 & \oai & \sysname & \sysnamec
 & \oai & \sysname & \sysnamec
 & \oai & \sysname & \sysnamec
\\ \hline \\
smcalflow 
 & 0.64 & 0.90 & 0.90* 
 & 0.72 & 0.94 & 0.94*
 & 0.74 & 0.94 & 0.95*
 & 0.75 & 0.94 & 0.95*
\\
Bash 
 & 0.75 & 0.78 & ?? 
 & 0.84 & 0.88 &
 & 0.85 & 0.89 & ??
 & 0.86 & 0.89 &
\\
Conala 
 & 0.77 & 0.75 & ?? 
 & 0.87 & 0.85 &
 & 0.88 & 0.86 & ??
 & 0.88 & 0.86 &
\end{tabular}
\end{center}
\end{table}
\endignore}

\ignore{ 
\begin{table}[t]
\caption{NL2SMCalFlow Results Using Different Few-Shot Selection Techniques}
\label{table:nl2smcalflow}
\begin{center}
\begin{tabular}{lccccc}
& static & OAI NL & OAI Code & Oracle & \sysname
\\ \hline \\
Exact Match & 0.2 & 0.51 & 0.66 & 0.72 & ??
\\
Edit Similarity &  
\\
\end{tabular}
\end{center}
\end{table}
\endignore}

There are a few reasons why \ada embeddings do well on the \conala dataset. 
The first is that the target language is \python, and the pretrained embedding models are 
known to be fairly good at Python.
The second is that the NL descriptions of the \python code are of very high quality.
The \conala dataset contains ``cleaned up'' versions of the original NL annotation extracted
from \stackoverflow. These NL 
descriptions were fixed by humans whenever they were found to not describe the code accurately. 
We used these ``fixed'' versions in our experiments. 
Since the NL describes the code fairly accurately, so its embedding 
(computed by \ada embeddings) is naturally very close to the embedding of the associated 
\python code, and hence there is not much scope for our technique to improve it.
Finally, and most importantly, we realized after the experiments were completed that the 
mapping from NL to \python code in 
the \conala dataset is one-to-many. While it is true that our technique can work with one-to-many
relations, it does so only after we have prepared the dataset by collecting all possible corpus 
elements that correspond to the same query, and since we did not perform this preprocessing, our
approach struggles on the \conala dataset. 

In contrast to \python, \smcalflow is an unfamiliar target and the \ada model does not do as well
on embedding those programs. Furthermore, the train-test split in the \smcalflow benchmark is such
that the train set contains good representatives for the test queries, and hence, our adaptation
performs extremely well on it.  

\section{Related Work}

End-to-end Information Retrieval (IR) systems colloquially have four major components~\citep{llm4ir}, namely (a) query rewriter, (b) retriever, (c) re-ranker and (d) reader. Our proposed technique \sysname potentially impacts the first three components, either directly or indirectly, by abstracting out the input representation and directly tuning the meta-representations (embeddings). Embeddings are at the core Neural IR. Particularly, recent advances in language understanding demonstrate how off-the-shelf LLMs and their embedding representations~\citep{openaiemb, touvron2023llama} are being leveraged in both upstream and downstream components of the IR pipeline \citep{dodge2020fine, asai2019learning, khashabi-etal-2020-unifiedqa}. This motivates our works on improving retrieval and its applications with minimal augmentations to off-the-shelf embeddings.

\subsubsection*{Query Rewrite}
In the IR domain, query formulation and expansion have been considered as essential steps to improve upstream retrieval. Classic techniques, such as pseudo-relevance feedback (PRF)~\citep{salton1990improving} and Rocchio's relevance feedback~\citep{rocchio1971relevance}, have been used to refine initial queries based on feedback from relevant documents. Moreover, external sources, such as thesauri and ontologies, have been employed to expand queries~\citep{aqe}. These methods have primarily operated in the textual space by manipulating query terms, adding synonyms or additional metadata to improve the robustness of the retrieval.

Our work potentially achieves the goals of query expansion, but using a significantly different approach. Rather than work on the input (textual) domain, we focus on the embedding space. We learn adapters that automatically perform query embedding refinement~\citep{aqe}. The refined embedding can be viewed as an embedding of the (unknown) rewritten query.  Our approach leverages the inherent structure and semantics captured in embeddings, and thus allows us to adapt queries in a data-driven, task-specific fashion, making our approach particularly effective in scenarios where queries and corpus elements involve diverse types of entities, such as NL-code and query-arguments, with a strict retrieval policy requirement.

\subsubsection*{Retriever}
The retrieval component of the pipeline varies widely according to the downstream task associated with the system. Here, the query and the search space could be either homogeneous~\citep{voorhees1999natural}, that is, the same information modalities, such as NL-NL, or heterogeneous~\citep{thakur2021beir}, that is, different but dependent information modalities, such as NL-Code. While building retrieval systems for dissimilar model types has always been a challenging problem, the use cases of such retrievals have been gathering momentum due to proliferation of LLM-based tools. Various approaches have been proposed for heterogenous retrieval, including learning embeddings in a joint cross-modal fashion by sharing semantic information in a common representational space within the vision-language community for cross-modal alignment and recommendation systems~\citep{ma2021joint}.

Joint Embedding Predictive Architectures (JEPAs)~\citep{assran2023selfsupervised} extend this idea further by not only mapping data into a shared representational space, but also learning predictive functions that can identify the compatibility or similarity between inputs. This predictive capability, besides being useful to act over user's history to present recommendations, can also be used to learn and suggest entities compatible with their preferences~\citep{assran2023selfsupervised}. Our technique draws inspiration from such architectures where we train \sysname with the loss function in the (possible) shared latent embedding space for heterogeneous entity pairs and optimizing the similarity in this representation. This space is tuned to emphasize the aspects of the representation that are learned from the data for a given downstream task. It can then be used as the ground for making multiple observations about the closeness between these dissimilar entities.

\subsubsection*{Reranker}
The Reranker aims to reorder the retrieved entities so that the top entities are more aligned with the query and task at hand. It involves sophisticated methods of ranking that consider more subtle features of the retrieved entities, such as term-proximity, quality, and relevance feedback~\citep{adaptiveRF,valcarce2018robustness,wang2009portfolio}. Neural models have been increasingly employed in learning to re-rank for a particular task setting~\citep{learn2rank,thakur2021beir,cormack2011efficient,dumais2003stuff,huang2009bayesian}.
\sysname enables developers to try out various types and combinations of loss functions, and tune the adaptation to align with their preferred ranking suited to their tasks. We do this with modest data pre-requisites and without learning complicated models that represent the ranking space.

\subsubsection*{Reader}
In the context of the emerging paradigm of Retrieval Augmented Generation (RAG)~\citep{lewis2020retrieval}, the Reader component plays a crucial role by synthesizing relevant information from retrieved information entities. 
RAG combines traditional information retrieval with natural language generation, often involving large language models, such as GPT~\citep{gpt, gpt2, gpt3, openai2023gpt4}, BERT~\citep{devlin2018bert, akkalyoncu-yilmaz-etal-2019-applying, bertRetrievalTuning} and T5~\citep{T5}; some of which are tuned on retrieval tasks~\citep{FlanT5, asai2022taskaware}. 
While \sysname does not directly help with the goals of the reader component, it can potentially empower these newer applications, such as RAG and few-shot selection, through its role in the other three retrieval components.








\section{Conclusion}

We presented adapted dense retrieval -- a dense retrieval approach that additionally adapts
the embedding and aligns it with any task-specific notion of similarity or closeness.
The specific adapter we consider in this paper adds a residual term to the general-purpose
embedding. We considered a very simple adapter that maintains a key-value dictionary and
computes the residual by softmax-based lookup in this dictionary. 
The benefit of using a small adapter is that it can be trained
using only a limited amount of annotated task-specific data. 
In future work, we plan to 
consider richer architectures for the adapter. We experimented with two kinds of adapter
architectures: one where we adapt only the query embedding and 
the other where we adapt both the query and corpus embeddings.

Just as fine tuning of generative models is only needed in special cases, adapters for
embeddings are necessary only when the semantic notion of similarity induced by default embeddings 
is not aligned with the task-specific notion.
This usually happens when the query or corpus elements are
uncommon entities that are unfamiliar to pretrained models, such as code in a niche
domain-specific language. 




%

\ignore{
\centerline{\bf Numbers and Arrays}
\bgroup
\def\arraystretch{1.5}
\begin{tabular}{p{1in}p{3.25in}}
$\displaystyle a$ & A scalar (integer or real)\\
$\displaystyle \va$ & A vector\\
$\displaystyle \mA$ & A matrix\\
$\displaystyle \tA$ & A tensor\\
$\displaystyle \mI_n$ & Identity matrix with $n$ rows and $n$ columns\\
$\displaystyle \mI$ & Identity matrix with dimensionality implied by context\\
$\displaystyle \ve^{(i)}$ & Standard basis vector $[0,\dots,0,1,0,\dots,0]$ with a 1 at position $i$\\
$\displaystyle \text{diag}(\va)$ & A square, diagonal matrix with diagonal entries given by $\va$\\
$\displaystyle \ra$ & A scalar random variable\\
$\displaystyle \rva$ & A vector-valued random variable\\
$\displaystyle \rmA$ & A matrix-valued random variable\\
\end{tabular}
\egroup
\vspace{0.25cm}

\centerline{\bf Sets and Graphs}
\bgroup
\def\arraystretch{1.5}

\begin{tabular}{p{1.25in}p{3.25in}}
$\displaystyle \sA$ & A set\\
$\displaystyle \R$ & The set of real numbers \\
$\displaystyle \{0, 1\}$ & The set containing 0 and 1 \\
$\displaystyle \{0, 1, \dots, n \}$ & The set of all integers between $0$ and $n$\\
$\displaystyle [a, b]$ & The real interval including $a$ and $b$\\
$\displaystyle (a, b]$ & The real interval excluding $a$ but including $b$\\
$\displaystyle \sA \backslash \sB$ & Set subtraction, i.e., the set containing the elements of $\sA$ that are not in $\sB$\\
$\displaystyle \gG$ & A graph\\
$\displaystyle \parents_\gG(\ervx_i)$ & The parents of $\ervx_i$ in $\gG$
\end{tabular}
\vspace{0.25cm}

\centerline{\bf Indexing}
\bgroup
\def\arraystretch{1.5}

\begin{tabular}{p{1.25in}p{3.25in}}
$\displaystyle \eva_i$ & Element $i$ of vector $\va$, with indexing starting at 1 \\
$\displaystyle \eva_{-i}$ & All elements of vector $\va$ except for element $i$ \\
$\displaystyle \emA_{i,j}$ & Element $i, j$ of matrix $\mA$ \\
$\displaystyle \mA_{i, :}$ & Row $i$ of matrix $\mA$ \\
$\displaystyle \mA_{:, i}$ & Column $i$ of matrix $\mA$ \\
$\displaystyle \etA_{i, j, k}$ & Element $(i, j, k)$ of a 3-D tensor $\tA$\\
$\displaystyle \tA_{:, :, i}$ & 2-D slice of a 3-D tensor\\
$\displaystyle \erva_i$ & Element $i$ of the random vector $\rva$ \\
\end{tabular}
\egroup
\vspace{0.25cm}

\centerline{\bf Calculus}
\bgroup
\def\arraystretch{1.5}
\begin{tabular}{p{1.25in}p{3.25in}}
$\displaystyle\frac{d y} {d x}$ & Derivative of $y$ with respect to $x$\\ [2ex]
$\displaystyle \frac{\partial y} {\partial x} $ & Partial derivative of $y$ with respect to $x$ \\
$\displaystyle \nabla_\vx y $ & Gradient of $y$ with respect to $\vx$ \\
$\displaystyle \nabla_\mX y $ & Matrix derivatives of $y$ with respect to $\mX$ \\
$\displaystyle \nabla_\tX y $ & Tensor containing derivatives of $y$ with respect to $\tX$ \\
$\displaystyle \frac{\partial f}{\partial \vx} $ & Jacobian matrix $\mJ \in \R^{m\times n}$ of $f: \R^n \rightarrow \R^m$\\
$\displaystyle \nabla_\vx^2 f(\vx)\text{ or }\mH( f)(\vx)$ & The Hessian matrix of $f$ at input point $\vx$\\
$\displaystyle \int f(\vx) d\vx $ & Definite integral over the entire domain of $\vx$ \\
$\displaystyle \int_\sS f(\vx) d\vx$ & Definite integral with respect to $\vx$ over the set $\sS$ \\
\end{tabular}
\egroup
\vspace{0.25cm}

\centerline{\bf Probability and Information Theory}
\bgroup
\def\arraystretch{1.5}
\begin{tabular}{p{1.25in}p{3.25in}}
$\displaystyle P(\ra)$ & A probability distribution over a discrete variable\\
$\displaystyle p(\ra)$ & A probability distribution over a continuous variable, or over
a variable whose type has not been specified\\
$\displaystyle \ra \sim P$ & Random variable $\ra$ has distribution $P$\\
$\displaystyle  \E_{\rx\sim P} [ f(x) ]\text{ or } \E f(x)$ & Expectation of $f(x)$ with respect to $P(\rx)$ \\
$\displaystyle \Var(f(x)) $ &  Variance of $f(x)$ under $P(\rx)$ \\
$\displaystyle \Cov(f(x),g(x)) $ & Covariance of $f(x)$ and $g(x)$ under $P(\rx)$\\
$\displaystyle H(\rx) $ & Shannon entropy of the random variable $\rx$\\
$\displaystyle \KL ( P \Vert Q ) $ & Kullback-Leibler divergence of P and Q \\
$\displaystyle \mathcal{N} ( \vx ; \vmu , \mSigma)$ & Gaussian distribution %
over $\vx$ with mean $\vmu$ and covariance $\mSigma$ \\
\end{tabular}
\egroup
\vspace{0.25cm}

\centerline{\bf Functions}
\bgroup
\def\arraystretch{1.5}
\begin{tabular}{p{1.25in}p{3.25in}}
$\displaystyle f: \sA \rightarrow \sB$ & The function $f$ with domain $\sA$ and range $\sB$\\
$\displaystyle f \circ g $ & Composition of the functions $f$ and $g$ \\
  $\displaystyle f(\vx ; \vtheta) $ & A function of $\vx$ parametrized by $\vtheta$.
  (Sometimes we write $f(\vx)$ and omit the argument $\vtheta$ to lighten notation) \\
$\displaystyle \log x$ & Natural logarithm of $x$ \\
$\displaystyle \sigma(x)$ & Logistic sigmoid, $\displaystyle \frac{1} {1 + \exp(-x)}$ \\
$\displaystyle \zeta(x)$ & Softplus, $\log(1 + \exp(x))$ \\
$\displaystyle || \vx ||_p $ & $\normlp$ norm of $\vx$ \\
$\displaystyle || \vx || $ & $\normltwo$ norm of $\vx$ \\
$\displaystyle x^+$ & Positive part of $x$, i.e., $\max(0,x)$\\
$\displaystyle \1_\mathrm{condition}$ & is 1 if the condition is true, 0 otherwise\\
\end{tabular}
\egroup
\vspace{0.25cm}

\endignore}


\bibliography{references}
\bibliographystyle{iclr2024_conference}


\end{document}